\documentclass[prl,aps,preprint,showpacs]{revtex4}
\usepackage{graphicx}

\usepackage{epstopdf}
\begin{document}

\title{Self-Consistent Model of Roton Cluster Excitations in Liquid Helium II}
\author{V. I. Kruglov and M. J. Collett}
\address{Physics Department, The University of Auckland, Private Bag
         92019, Auckland, New Zealand}

\begin{abstract}

We have proposed a model of roton cluster excitations in liquid helium~II based on a Schr\"odinger-type equation with a self-consistent confining potential. We have derived an equation for the number of atoms in roton excitations, which can be treated as quantum $3{\rm D}$ solitons, depending on vibrational quantum numbers. It is shown that the smallest roton cluster is in the symmetric vibrational quantum state and consists of 13 helium atoms. We have also used a modified Born approximation to calculate the $s$-scattering length for helium atoms. This allows us to calculate all parameters of Landau's roton excitation spectrum, in agreement to high accuracy with experimental measurements from neutron scattering.

\end{abstract}


\maketitle

\section{1 introduction}

Liquid helium is a unique system of strongly interacting atoms which becomes a quantum liquid (helium~II) when the temperature is lower then some critical temperature $T_c$. This critical temperature $T_c$ can be well estimated by two parameters:  the de Broglie thermal  wavelength  $\lambda_D=2\pi\hbar(3mk_{\mathrm B} T)^{-1/2}$ and the $s$-scattering length $a_0= 2.2$~\AA~of the helium atoms. Thus we may suppose that in liquid helium, a phase transition occurs to a quantum liquid when the de Broglie thermal  wavelength overlaps two nearest neighbors of helium atoms: $\lambda_D= 2d$, where $d=2a_0$ is the effective diameter of helium atoms. This equation yields the temperature $T_c= 2.05~{\rm K}$, which is very close to the experimental value $T_c=2.18~{\rm K}$ of the phase $\lambda$-transition in liquid helium. 

The study of liquid helium~II, and  quantum liquids more generally, has been an active area of experimental and theoretical research ever since Kapitza's \cite{Kap} discovery of the superfluidity in helium~II. 
Based on some experimental facts, Landau~\cite{Land} assumed that in liquid helium~II there exist two types of elementary excitations: phonons and rotons. He also phenomenologically introduced the roton energy spectrum (see Eq.(\ref{1})) to explain the thermodynamic behavior of liquid helium~II.
Bogoliubov~\cite{Bogoliubov-47} first found theoretically the spectrum of quasi-particles in a dilute Bose gas. This theory can be used for a Bose gas when the temperature is close to zero, the particle density  $n$ is low ($\sqrt{a_0^3n}\ll1$),  and the wavenumbers of the particles are sufficiently small ($k a_0 \ll1$).
The relation between the
energy spectrum of the elementary excitations of liquid $^4$He and the structure
factor of the liquid was found by Feynman~\cite{Feyn}. Feynman also
proposed a phenomenological model of the roton as a localized
vortex ring with characteristic size of the order of the mean atomic
distance in liquid helium~II. The vortex rings were observed experimentally by Rayfield and Reif \cite{Rey}; however, there has been no experimental confirmation that the vortex ring and roton are the same excitations in liquid helium. 

We also note that a variety of many-body theories have been applied to the problem since 1990 and a quantitative description of a roton has been obtained from variational integral equations, from a variational Monte Carlo method and by an imaginary time exact path integral.
Thus, a correct model of roton quasi-particles should be based on ``first principles'' and lead to quantitative agreement with experimental data. Recently such a model was proposed in the paper~\cite{Krug}, where it was found that the roton is an
excitation of a cluster of $13$ atoms, presumably having a central atom surrounded by a shell of $12$ atoms situated at the vertices of a regular icosahedron. 
In that paper we considered the number of atoms $N_c$ of the roton cluster as a fitting parameter that was found from the experimental data of Yarnell {\it et al}~\cite{Yar}.

In fact, the theory developed in \cite{Krug} does not need fitting to experimental data. To formulate this theory completely self-consistently we should find an additional equation for the chemical potential of roton excitations. In this paper, we calculate the vibrational energy spectrum of the roton cluster excitations with an arbitrary number of atoms in the cluster by the Schr\"odinger-type equation with a self-consistent confining potential derived in \cite{Krug}.
Using scattering theory we have formulated the modified Born approximation (MBA) and have found the $s$-scattering length $a_0$ for helium atoms from the Lennard-Jones interatomic potential. 
We also derive in this paper the equation for the chemical potential $\mu$ of a roton cluster, which allows us to show that the smallest stable roton clusters in liquid helium~II indeed contain $13$ atoms. Moreover, we have found the numbers of atoms $N_c$ for the roton clusters which are in a vibrational state $|n_1n_2n_3\rangle$ with arbitrary quantum numbers $n_1,n_2,n_3$.  
This self-consistent approach allows us to accomplish all calculations for the parameters of the roton cluter excitations in helium~II and leads to a quantitative agreement with experimental data.
In particular, the roton clusters with $N_c=13$ are in the symmetric vibrational state $|001\rangle$ and the energy spectrum of roton cluster excitations can be derived in the form~\cite{Krug} :
\begin{equation}
\varepsilon(p) = \Delta + \frac{(p-p_0)^2}{2\mu_{\mathrm R}}\ ,
\label{1}
\end{equation}
which coincides with Landau's empirical formula.

\section{2  Schr\"odinger equation for roton excitations}

The nonlinear Schr\"odinger equation (NLSE) for roton cluster excitations can be derived by a Hartree-Fock variation procedure for the trial wave function $\Psi_R({\bf x}_1,...,{\bf x}_n;t)=C\psi_R({\bf x}_1,t)...\psi_R({\bf x}_n,t)$ where $n=N_c$. In this procedure we represent the interaction potential by two parts: $U_R({\bf x})$ and $U_A({\bf x})$, where $U_R({\bf x})$ is the repulsive part and $U_A({\bf x})$ is the attractive part.
The repulsive part of this potential can be approximated by a hard sphere potential,
\begin{equation}
U_R({\bf x}-{\bf x}')=\frac{4\pi a_0\hbar^2}{m}\delta({\bf x}-{\bf x}'),
\label{eR}
\end{equation}
and the attractive part (the confining many-body effective potential) is  
\begin{equation}
U_A({\bf x})=(m/2) \sum_{k=1}^3 \omega_k^2 x_k^2.
\label{eA}
\end{equation}
The parameters $\omega_k^2$ are found by the stationarity conditions
$(\partial V/\partial a_k)_{a_k=\bar{a}_k}=0$ (see below).
This potential takes into account all forces between $^4{\rm  He}$ atoms of the cluster and those of the surroundings, and also long-range many-body attractive forces among atoms of the cluster~\cite{Krug}. This Hartree-Fock variation procedure directly leads to a NLSE : 
\begin{equation}
i\hbar\frac{\partial\psi_R}{\partial t}=\left(-\frac{\hbar^2}{2m}\nabla^{2}+\frac{m}{2}
\sum_{k=1}^{3}\omega_{k}^{2}x_{k}^{2}
+\frac{4\pi a_{0}\hbar^2 (N_c-1)}{mN_c}|\psi_R|^{2}
\right)\psi_R.
\label{e1}
\end{equation}
Here the normalisation condition for the wave function $\psi_R({\bf x},t)$ has the form $\int_{V_c}|\psi_R|^{2}d^3{\bf x}=N_c$ where $V_c=N_c/n$ is the roton cluster volume and the parameters $\omega_k^2$ are given by Eq.(\ref{3}). 
Because Eq.(\ref{e1}) follows from a Hartree-Fock variational procedure the number of atoms forming the roton cluster should satisfy the condition $N_c>1$ (not necessarily $N_c\gg1$). But below we will suppose the spherical symmetry of the roton clusters, which can only be accurate with a large number of atoms (for example one may demand $N_c\geq 10$).

Despite its apparently similar form,  equation (\ref{e1}) should not be confused with the well-known Gross-Pitaevskii equation~\cite{Pit,Gros}.   Both contain the same hard-sphere repulsive potential to describe short-range collisional interactions.  However, in the Gross-Pitaevski equation the attractive harmonic potential, if present, describes an external trap; in  equation (\ref{e1}) the harmonic $U_A({\bf x})$ is a effective  long-range many-body interaction, found by the variational procedure.  Thus while the Gross-Pitaevski equation is only valid in the limit of low particle density, $a_0^3n\ll1$, and small wavenumbers, $k a_0 \ll1$, equation (\ref{e1}) remains valid at high densities, $a_0^3n\simeq 1$, provided $|k-k_0|a_0\ll1$ (or more precisely, provided $|k-k_0|\leq 0.2~\AA^{-1}$). In the limit when $N_c\rightarrow \infty$ and $\omega_k^2\rightarrow 0$ the confining many-body potential $U_A({\bf x})$ vanishes and equation (\ref{e1}) does becomes just the Gross-Pitaevskii equation. 

Equation (\ref{e1}) is semiclassical, and as mentioned above, all phenomena in helium II, including roton excitations, should be described quantum mechanically.  We present below a quantisation procedure for the NLSE based on a self-similar solution.

Following the method of~\cite{Krug2} (for the $3D$ case see~\cite{Krug3}) it can be shown that this NLSE (\ref{e1}) has a self-similar solution when the dimensionless parameter $\varepsilon_c=N_c^{1/3}/[6(N_c-1)]$ satisfies the condition $\varepsilon_c\ll 1$. Evidently, this is true for $N_c\geq10$  because in this case $\varepsilon_c\leq 0.04$, and even for $N_c=3$ we find that this parameter is small enough: $\varepsilon_c=0.12$. Further, we have already shown~\cite{Krug} that the self-similar evolution of the atoms in the roton clusters described by the NLSE (\ref{e1}) can be derived from a  Hamiltonian in the space of the ellipsoidal coordinates  of the roton cluster: 
\begin{equation}
 {\mathcal H} = \sum_{k=1}^3 \frac{\pi_k^2}{2m_0}
 + \frac{m_0}{2} \sum_{k=1}^3 \omega_k^2 a_k^2
 + \frac{G}{a_1 a_2 a_3}\ ,
 \label{2}
\end{equation}
where $\pi_k=m_0 da_k/dt$ are the canonical momenta, $a_k$ are the ellipsoidal coordinates, and $m_0=m/7$ is the renormalized mass. The interaction constant $G$ and the parameters $\omega_k^2$ of the self-consistent confining potential $U_R(a_1,a_2,a_3)=(m_0/2) \sum_{k=1}^3 \omega_k^2 a_k^2$ of the roton cluster in Eq.(\ref{2}) are given by~\cite{Krug} : 
\begin{equation}
\omega_k^2 = \frac{15(N_{\mathrm
c}-1)a_0\hbar^2}{m^2\bar{a}_1\bar{a}_2\bar{a}_3\bar{a}_k^2},
~~~G = \frac{15(N_{\mathrm c}-1)a_0\hbar^2}{7m}\ .
\label{3} 
\end{equation}
 The quantisation of the Hamiltonian~(\ref{2}) yields a Schr\"odinger equation
for the roton wave function $\chi(a_1,a_2,a_3)$~:
\begin{equation}
{\mathrm i}\hbar \frac{\partial\chi}{\partial t} =
\left( - \frac{\hbar^2}{2\,m_0}
\sum_{k=1}^3 \frac{\partial^2}{\partial a_k^2}
+ V(a_1,a_2,a_3) \right) \chi\,,
\label{4}
\end{equation}
where the potential $V(a_1,a_2,a_3)$ has the form
\begin{equation}
V(a_1,a_2,a_3) = \frac{m_0}{2} \sum_{k=1}^3 \omega_k^2 a_k^2
+ \frac{G}{a_1 a_2 a_3}\,.
\label{5}
\end{equation}
The parameters $\omega_k^2$ must satisfy the stationarity conditions
$(\partial V/\partial a_k)_{a_k=\bar{a}_k}=0$, which just yield
Eq.(\ref{3}) again. We note that the equation (\ref{4}) is not the standard Schr\"odinger equation because it given in the space of ellipsoidal parameters $a_k$. This  Schr\"odinger type equation can be solved by an expansion of the potential energy (\ref{5}) in a series with the variable $y_k=a_k-\bar{a}_k$, subject to the condition $|y_k|/\bar{a}_k\ll1$. In the quadratic approximation we find 
\begin{equation}
V(a_1,a_2,a_3)=V_{0}+\frac{m_{0}}{2}\sum_{k=1}^{3}\sum_{s=1}^{3}\Lambda_{ks}y_{k}y_{s},
\label{6} 
\end{equation}
where the constant $V_{0}$ and the matrix $\Lambda_{ks}$ are given by
\begin{equation}
V_{0}=\frac{5G}{2\overline{a}_{1}
\overline{a}_{2}\overline{a}_{3}},~~~\Lambda_{ks}=\omega_{k}\omega_{s}+2\omega_{k}^{2}\delta_{ks}~.
\label{7}  
\end{equation}
The eigenvectors $u_{s}^{(k)}$ of the matrix $\Lambda_{ks}$ obey 
\begin{equation}
\sum_{s=1}^{3}\Lambda_{is}u_{s}^{(k)}=\Omega_{k}^{2}u_{i}^{(k)},~~~\sum_{s=1}^{3}u_{s}^{(i)}u_{s}^{(k)}=\delta_{ik},
\label{8}
\end{equation}
where the second equation is just the normalisation condition. Thus the coordinate transformation
\begin{equation}
y_{s}=a_s-\bar{a}_s=\sum_{k=1}^{3}u_{s}^{(k)}z_{k},
\label{9}
\end{equation}
is orthogonal. Using Eqs.(\ref{6}-\ref{9}), we can now write the Schr\"odinger equation (\ref{5}) as
\begin{equation}
i\hbar\frac{\partial}{\partial t}\chi=\left(-\frac{\hbar^{2}}{2m_{0}}
\sum_{k=1}^{3}\frac{\partial^{2}}{\partial z_{k}^{2}} 
+\frac{m_{0}}{2}\sum_{k=1}^{3}\Omega_{k}^{2}z_{k}^{2}+ V_{0}\right)\chi.
\label{10}
\end{equation}
 We note that the energy spectrum of Eq.(\ref{10}) does not depend on the renormalized mass $m_0$ (which for some problems can be redefined ~\cite{Krug3}).
Here the eigenfrequencies $\Omega_{k}$ are given by equation $\det (\Lambda-\Omega_k^2{\rm I})=0$ where ${\rm I}$ is the identity matrix, or, in explicit form,
\begin{eqnarray}
& &\left(\Omega_{k}^{2}-3\omega_{1}^{2}\right)\left(\Omega_{k}^{2}-3\omega_{2}^{2}\right)\left(\Omega_{k}^{2}-3\omega_{3}^{2}\right)
-\omega_{2}^{2}\omega_{3}^{2}\left(\Omega_{k}^{2}-3\omega_{1}^{2}\right)
\nonumber\\
& &-\omega_{1}^{2}\omega_{3}^{2}\left(\Omega_{k}^{2}-3\omega_{2}^{2}\right)
-\omega_{1}^{2}\omega_{2}^{2}\left(\Omega_{k}^{2}-3\omega_{3}^{2}\right)
-2\omega_{1}^{2}\omega_{2}^{2}\omega_{3}^{2}=0 ~.
\label{11}
\end{eqnarray}
This allows us to find the eigenfunctions $\chi_{n_{1}n_{2}n_{3}}$ and eigenenergies $E_{n_{1}n_{2}n_{3}}$ by 
\begin{equation}
\hat{{\mathcal H}}\chi_{n_{1}n_{2}n_{3}}=E_{n_{1}n_{2}n_{3}}\chi_{n_{1}n_{2}n_{3}},
\label{12}
\end{equation}
where the Hamiltonian $\hat{{\mathcal H}}$ is that defined by the Schr\"odinger equation (\ref{10}). These eigenfunctions $\chi_{n_{1}n_{2}n_{3}}$ are 
\begin{equation}
\chi_{n_{1}n_{2}n_{3}}(z_1,z_2,z_3)=\left(\frac{m_{a}^{3}
\Omega_{1}\Omega_{2}\Omega_{3}}{\hbar^{3}\pi^{3}}\right)^{1/4}\left(
\frac{2^{-n_{1}-n_{2}-n_{3}}}{n_{1}!n_{2}!n_{3}!}\right)^{1/2}
\prod_{k=1}^{3}H_{n_{k}}(\xi_{k})\exp(-\frac{1}{2}\xi_{k}^{2}),
\label{13}
\end{equation} 
where $H_{n}(\xi)$ are Hermite polynomials and $\xi_{k}=z_{k}\sqrt{m_0\Omega_k/\hbar}$.
The eigenenergies in Eq.(\ref{12}) have the form
\begin{equation}
E_{n_{1}n_{2}n_{3}}=V_{0}+\hbar\sum_{k=1}^{3}\Omega_{k}(n_{k}+1/2),
\label{14} 
\end{equation}
where $n_k=0,1,2,...$ at $k=1,2,3$ are the quantum numbers of a 3D quantum harmonic oscillator.
Assuming now spherical symmetry of the roton clusters, we find
$\bar{a}_k=\bar{a}$ (for $k=1,2,3$) where $\bar{a}$ is given by ~\cite{Krug} :
\begin{equation}
\bar{a}=q_0N_c^{1/3},~~~q_0=\left(\frac{3}{4\pi n}\right)^{1/3}.
\label{15}
\end{equation}
Here $n=\rho/m$ is the particle density of liquid helium, giving $q_0= 2.22$~\AA. For example, in the case $N_c=13$ we have $\bar{a}= 5.22$~\AA. We note that the value of the parameter $q_0$ defined by the particle density $n$ of liquid helium is very close to the value of the $s$-scattering length $a_0= 2.2$~\AA~for the helium atoms. From Eq.(\ref{15}) it also follows that $\bar{a}=q_0$ when $N_c=1$, so the physical sense of the parameter $q_0$ is the same as that of $a_0$: an effective radius of the helium atoms.

In this spherically symmetric case Eq.(\ref{11}) yields the eigenfrequencies 
$\Omega_1=\Omega_2=\sqrt{2}\omega$, $\Omega_3=\sqrt{5}\omega$. Here the frequency $\omega=[15 (N_c-1) a_0]^{1/2}\hbar m^{-1}\bar{a}^{-5/2}$ follows from Eq.(\ref{3}) when $\bar{a}_k=\bar{a}$ for $(k=1,2,3)$. Let us define the energy of a symmetric excitation $\Delta_s=\hbar\Omega_3$ and the energy of an asymmetric excitation $\Delta_a=\hbar\Omega_1=\hbar\Omega_2$.  Then, taking into account Eq.(\ref{15}), we find 
\begin{equation}
\Delta_s=5\sqrt{3}\left(\frac{a_0}{q_0}\right)^{5/2}\frac{(N_c-1)^{1/2}}{N_c^{5/6}}\frac{\hbar^2}{ma_0^2},~~~\Delta_a=\sqrt{2/5}\Delta_s.
\label{16}
\end{equation}

\section{3 roton gap}

At first we consider in detail the calculation of the $s$-scattering length $a_0$ for helium atoms based on the Lennard-Jones intermolecular potential. The Schr\"odinger equation for the $s$-scattering wave and appropriate boundary conditions are : 
 \begin{equation}
\phi''(r)+k^2\phi(r)=\frac{2m_c}{\hbar}U(r)\phi(r),
\label{17}
\end{equation}
\begin{equation}
\phi(0)=0,~~~\phi(r)\rightarrow \frac{1}{k}\sin(kr+\delta_0(k))~~\mbox{as}~~r\rightarrow \infty,
\label{18}
\end{equation} 
where $k=p/\hbar$ is the wavenumber, $\phi(r)$ is the radial $s$-scattering wavefunction, $\delta_0(k)$ is  the phase shift of the $s$-scattering wave, $m_c$ is the reduced mass, and $U(r)$ is the scattering potential. Defining the $s$-wave scattering length as usual by $a_0^{-1}=-\lim_{k\rightarrow 0}[k {\rm cotg}~\delta_0(k)]$, we can write it in the form~\cite{Ta} :  
\begin{equation}
a_0=\lim_{k\rightarrow 0}~\frac{2m_c}{\hbar^2}\int_0^\infty U(r)\phi(r)rdr,
\label{19}
\end{equation}
(with the appropriate normalisation condition for the wave function).  
When the Born approximation is valid one can make in Eq.(\ref{19}) the substitution $\phi(r)\rightarrow \phi_0(r)$ where $\phi_0(r)= k^{-1}\sin(kr)$ is the wave function of free particle. However, for slow particles this approximation is generally meaningless~\cite{Ta} because of divergence of the integral in Eq.(\ref{19}). In the case $a_0>0$ we can instead formulate a modified Born approximation (MBA),  based on the substitution $\phi(r)\rightarrow \theta(r-a_0)k^{-1}\sin(kr)$ in Eq.(\ref{19}), where $\theta(r)$ is the Heaviside step function. Thus we suppose that the wave function is zero when $r<a_0$ (i.e. the region $r<a_0$ is unattainable
for a slow particle).  This leads to the equation :
\begin{equation}
a_0=\frac{2m_c}{\hbar^2}\int_{a_0}^\infty U(r) r^2dr.
\label{20}
\end{equation} 
For example, let us consider intermolecular interactions with the potential $U(r)=\alpha r^{-n}$ where $\alpha>0, n>3$. Eq.(\ref{20}) then yields
\begin{equation}
a_0=C_n\left(\frac{2m_c\alpha}{\hbar^2}\right)^{1/(n-2)},~~~C_n=\left(\frac{1}{n-3}\right)^{1/(n-2)}.
\label{21}
\end{equation}
The exact solution of Eqs.(\ref{17},\ref{18}) for this potential leads to an equation which differs from Eq.(\ref{21}) only in the coefficient: $C_n=(n-2)^{-2/(n-2)}\Gamma(\frac{n-3}{n-2})/\Gamma(\frac{n-1}{n-2})$, where $\Gamma(x)$ is the gamma function. Note that in the Born approximation the integral diverges in Eq.(\ref{19}). For example, at $n=6$ we find $C_6= 0.76$ in the MBA and $C_6= 0.68$ in the exact solution; at $n=12$ we have $C_{12}= 0.79$ in the MBA and $C_{12}= 0.71$ in the exact solution. 

We note that the MBA leads to the most accurate results when the potential $U(r)$ has some characteristic length $r_0$ where $a_0<r_0$. It asymptotically leads to the exact solutions for exactly integrable model potentials when $a_0\ll r_0$ . The condition $a_0<r_0$ does hold for helium intermolecular interactions given by the Lennard-Jones potential: $U(r)=4\epsilon[(r_0/r)^{12}-(r_0/r)^6]$. In this case calculating the integral in Eq.(\ref{20}) one can find a fifth order algebraic equation :
\begin{equation}
\rho^5-3\rho^2-\lambda=0,~~~\lambda=\frac{9\hbar^2}{8\epsilon m_cr_0^2},
\label{22}
\end{equation} 
where $\rho=(r_0/a_0)^2$ and $m_c=m/2$. We use in this paper  the parameters of the intermolecular Lennard-Jones potential for helium atoms calculated by a self-consistent-field Hartree-Fock method~\cite{Ahl}: $\epsilon/k_B=10.6~{\rm K}$ and $r_m=2.98$~\AA, where $r_m=2^{1/6}r_0$.  For these values we find from Eq.(\ref{22}) that $\rho= 1.468$, and $a_0=2.19$~\AA. Taking into account the limited accuracy of our calculations  we will assume in this paper  that $a_0= 2.2$~\AA  ~\cite{Krug,Pat}, which is very close to the result based on this MBA method.

Let us denote ${\cal E}_R$ the energy per particle of the roton cluster in the ground state and ${\cal E}_0$ the energy per particle of ground state of the bulk. Then taking into account that the energy of a bound state of two particles in the roton cluster is given by $\epsilon=-U(r_m)$ (to within some additive constant $\epsilon_b$) and the energy of two particles in the ground state of the bulk is $\epsilon_0=-U(2a_0)$ (to within an additive constant $\epsilon_b$) we can find
\begin{equation}
{\cal E}_R=-\epsilon-\epsilon_b,~~~{\cal E}_0=-\epsilon_0-\epsilon_b,
\label{23}
\end{equation} 
where $\epsilon_b$ is the contribution of the bulk to the ground state of the roton cluster. Here $U(r)$ is  the Lennard-Jones potential. 
Thus the chemical potential $\mu$ and the gap $\Delta$ of the roton cluster are~:
\begin{equation}
\mu= {\cal E}_R-{\cal E}_0=-\epsilon+\epsilon_0,~~~\Delta=\epsilon-\epsilon_0,
\label{24}
\end{equation} 
The equation $\Delta=\epsilon-\epsilon_0$ is fulfilled because from Eq.(\ref{1}) it follows that $\mu=-\Delta$. 
Using the above definitions one can find the positive parameter $\epsilon_0=-4\epsilon[(r_0/2a_0)^{12}-(r_0/2a_0)^6]$ which yields $\epsilon_0/k_B=1.95~{\rm K}$. Thus from Eq.(\ref{24}) follows that $\Delta/k_B= 8.65~{\rm K}$. This value coincides to three decimal places with the experimental value of the roton gap~\cite{Yar}: $\Delta/k_B=(8.65\pm 0.04)~{\rm K}$. 
Because ${\cal E}_0$ is the energy per particle of the ground state of the bulk we can write
\begin{equation} 
\lim_{N\rightarrow \infty}\frac{E_0}{N}={\cal E}_0=-\epsilon_0-\epsilon_b,
\label{25}
\end{equation}
where $E_0$ is the ground energy of the bulk and $N$ is the number of atoms in the bulk.
The ground state energy per particle ${\cal E}_0$ in the helium bulk can be found via the diffusion Monte Carlo method~\cite{Blum, Hamm}: ${\cal E}_0/k_B=-7.15~{\rm K}$. This yields the bulk parameter $\epsilon_b/k_B= 5.2~{\rm K}$. 

\section{4 roton cluster excitation numbers}

The equation $\Delta_s=-\mu$ holds only for the symmetric roton cluster excitation~\cite{Krug} with vibrational quantum numbers $n_1=n_2=0$ and $n_3=1$. In the general case, the roton cluster excitations with arbitrary vibrational quantum numbers satisfy the generalised equation for the number of atoms $N_a$  in the roton cluster: $E_{n_{1}n_{2}n_{3}}-E_{000}=-\mu$, where $E_{n_{1}n_{2}n_{3}}>E_{000}$. Hence, combining Eqs.(\ref{14}) and (\ref{24}) we find the generalised equation for the number of atoms $N_a$  in the roton cluster in the form
\begin{equation}
n_a\Delta_a+n_3\Delta_s = \epsilon-\epsilon_0,
\label{26}
\end{equation} 
\begin{figure}
\includegraphics[width=5in]{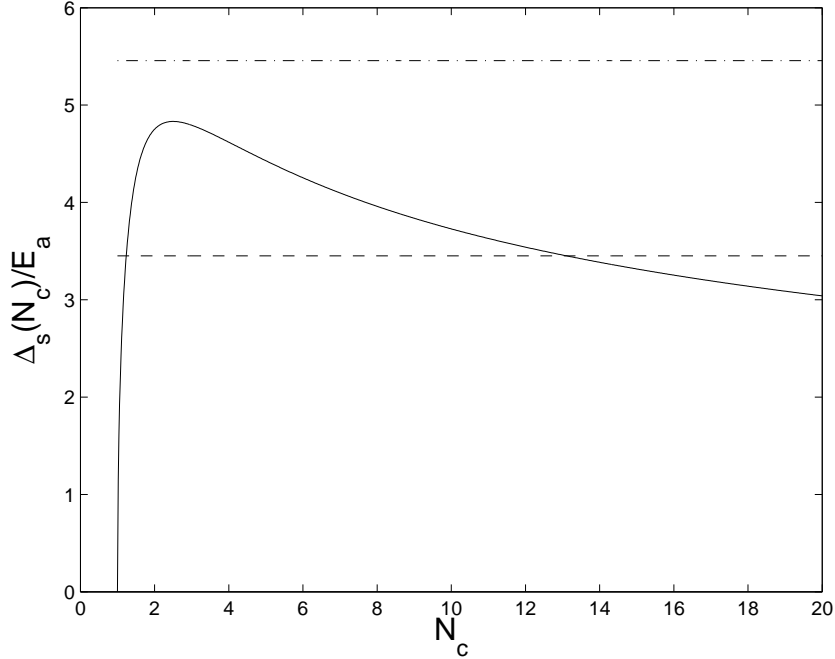}
\caption{The full curve is the dimensionless function $\Delta_s(N_c)/E_a$ (left side in (\ref{27})). The broken line is the constant $(\epsilon-\epsilon_0)/E_a$ (right side in (\ref{27}) at $n_a=0, n_3=1$), and the chain line is the constant $\sqrt{5/2}(\epsilon-\epsilon_0)/E_a$ (right side in (\ref{27}) at $n_a=1, n_3=0$). This figure shows that a stable ($d\Delta_s/dN_c<0$) solution of Eq.(\ref{27}) at $n_a=0, n_3=1$ is $N_c=N_{01}= 13.1\simeq13$, and there are no solutions of Eq.(\ref{27}) at $n_a=1, n_3=0$. It also shows that stable ($d\Delta_s/dN_c<0$) roton clusters are possible only when the integral numbers $N_c$ satisfy $N_c\geq 3$.}
\label{fig:1}
\end{figure}
where $n_a=n_1+n_2$ (with $n_1+n_2+n_3>0$), $\Delta_s=\hbar\Omega_3$, and  $\Delta_a=\hbar\Omega_1=\hbar\Omega_2$. Using Eq.(\ref{16}) we can write  Eq.(\ref{26}) in explicit form 
\begin{equation}
5\sqrt{3}\left(\frac{a_0}{q_0}\right)^{5/2}\frac{(N_c-1)^{1/2}}{N_c^{5/6}}=\frac{\epsilon-\epsilon_0}{(\sqrt{2/5}n_a+n_3)E_a},
\label{27}
\end{equation} 
where again $n_a+n_3>0$, and $E_a\equiv \hbar^2/(ma_0^2)$. Eq.(\ref{27}) is equivalent to a fifth order algebraic equation for the variable $N_c$. We can formulate the stability conditions of the roton cluster as $d(E_{n_1n_2n_3}-E_{000})/dN_c<0$ and $\lambda_D~\raisebox{-0.9ex}{$\stackrel{\textstyle{>}}{\sim}$}~2\bar{a}$, where we are considering here the number of atoms $N_c$ in the roton cluster as  a variable parameter. Combining these equations with Eq.(\ref{4}) and Eq.(\ref{15}) one can rewrite these two stability conditions in the form
\begin{equation}
\frac{d\Delta_s}{dN_c}<0,~~~\lambda_D~\raisebox{-0.9ex}{$\stackrel{\textstyle{>}}{\sim}$}~2q_0N_c^{1/3},
\label{28}
\end{equation} 
where $\Delta_s(N_c)$ is given by Eq.(\ref{16}) and $\lambda_D= 12.6 T^{-1/2}$\AA~is the thermal de Broglie wavelength of a helium atom. The numerical solutions of Eq.(\ref{27}) for the parameter  $N_c=N_{n_an_3}$  depending on quantum numbers $n_a=n_1+n_2$ and $n_3$ for a few lower-lying roton cluster excitations satisfying the the stability condition $d\Delta_s/dN_c<0$  are (numerical results in parentheses): $N_{01}=13~ (13.1); N_{20}= 28~ (28.3); N_{11}= 63~ (62.7); N_{30}= 99~ ( 99.3); N_{02}= 117~ (116.6); N_{21}= 170~ ( 170); N_{12}= 268~ (267.8); N_{03}= 397~ (397)$.  The cases $n_a=0, n_3=1$ and $n_a=1, n_3=0$  are shown in Fig. 1.

The second condition in (\ref{28}) yields $T~\raisebox{-0.9ex}{$\stackrel{\textstyle{<}}{\sim}$}~1.5~{\rm K}$ for $N_c=N_{01}=13$; $T~\raisebox{-0.9ex}{$\stackrel{\textstyle{<}}{\sim}$}~0.9~{\rm K}$ for $N_c=N_{20}=28$; and, for example, $T~\raisebox{-0.9ex}{$\stackrel{\textstyle{<}}{\sim}$}~0.3~{\rm K}$ for $N_c=N_{02}=117$. But it is known that the spectrum of phonon excitations in liquid helium~II  explains~\cite{Kha} the Debye law of the heat capacity $\sim T^3$ when $T~\raisebox{-0.9ex}{$\stackrel{\textstyle{<}}{\sim}$}~0.6~{\rm K}$. Assuming that the symmetric vibrational excitations of the roton clusters are more probable then asymmetric we find that the stable roton clusters in the region $0.6~\raisebox{-0.9ex}{$\stackrel{\textstyle{<}}{\sim}$}~T~\raisebox{-0.9ex}{$\stackrel{\textstyle{<}}{\sim}$}~1.5~{\rm K}$ in the main consist of $13$ helium atoms.
Eq.(\ref{26}) at $n_a=0$ and $n_3=1$ yields $\Delta_s=\Delta$ and hence $N_c=13$. 

\section{5 parameters of the roton excitation spectrum}

One can calculate the value $\Delta_s/k_B= 8.67~{\rm K}$ at $N_c=13$ which coincides within $\simeq 0.2\%$ with our theoretical value $\Delta/k_B= 8.65~{\rm K}$. The effective roton mass in Eq.(\ref{1}) is~\cite{Krug} $\mu_R=p_0^2/(4N_c\langle{\mathcal K}\rangle)$,
where  $\langle{\mathcal K}\rangle=\langle{\mathcal K_I}\rangle+\langle{\mathcal K_T}\rangle$ is the average energy of a particle in the roton cluster. Here the internal energy has the value ~\cite{Krug} $\langle{\mathcal K_I}\rangle/k_B=4.95~{\rm K}$ at $T=1.1~{\rm K}$ and $\langle{\mathcal K_T}\rangle/E_a=0.96(T/T_c)^{5/2}$ is the dimensionless energy of an ideal Bose gas per particle where $E_a=\hbar^2/(ma_0^2)$. Employing in this last formula the experimental value $T_c=2.18~{\rm K}$ gives $\langle{\mathcal K_T}\rangle/k_B=0.43~{\rm K}$ at $T=1.1~{\rm K}$. This yields $\mu_R/m=0.16$ for $N_c=13$ and $k_0=p_0/\hbar= 1.92$~\AA$^{-1}$, coinciding with the experimental value~\cite{Yar}. 

The momentum $p_0$ at the minimum of roton spectrum in Eq.(\ref{1}) can be approximately evaluated~\cite{Feyn} as $p_0=2\pi\hbar/{\mathcal R}$, where ${\mathcal R}$ is the average distance between atoms in the localized bound state. In the classical approximation one can assume that ${\mathcal R}=r_m$, so that $k_0=p_0/\hbar= 2.1 $~\AA$^{-1}$. In the quantum case, for symmetric vibrational excitations ($n_1=n_2=0, n_3=1$), Eq.(\ref{14}) yields $\xi_3=z_3/{\mathcal R}_q$ where ${\mathcal R}_q=\sqrt{\hbar/(m_0\Omega_3)}$, and from  Eq.(\ref{26}) we have that $\Delta_s=\Delta$. Hence the momentum $p_0$ at the minimum of the roton spectrum excitation has the form~:
\begin{equation}
p_0=\frac{2\pi\hbar}{{\mathcal R}_q}=2\pi\sqrt{m_0\Delta},
\label{29}
\end{equation}
where $m_0=m/7$ and $\Delta=\epsilon-\epsilon_0$. 
From this one can calculate $k_0=p_0/\hbar= 2$~\AA$^{-1}$,  about $4\%$ from the experimental value ~\cite{Yar,Hen}. 
One also can find a boundary point $k_c=p_c/\hbar$ in the elementary excitation spectrum between linear (phonon) and nonlinear regions as  $k_c=2\pi/d_c$, where $d_c=2\bar{a}= 10.44$~\AA~is the diameter of the roton cluster at $N_c=13$. Hence $k_c= 0.6$~\AA$^{-1}$, which again agrees with experiment~\cite{Yar,Hen}. We also note that the formfactor $S(k)$ for neutron scattering in liquid helium~II is connected with the energy spectrum of roton cluster excitations $\varepsilon(p)$ (see Eq.(\ref{1})) by Feynman's well-known approximate formula \cite{Fey}: $S(k)=\hbar^2k^2/(2m\varepsilon(\hbar k))$. Here $1.7 \AA^{-1}\leq k \leq2.1 \AA^{-1}$ because Eq.(\ref{1}) is valid only in the vicinity of the minimum of the roton spectrum excitation $k_0=p_0/\hbar$.

\section{6 conclusion}
 
In conclusion, we have proposed a full theoretical description of the model of roton cluster excitations in liquid helium~II. Our work shows the important role of the attractive forces in the dynamic behavior of the roton excitations in helium
that leads for some range of temperatures to the formation of  localised bound states or clusters which also can be treated as quantum 3D solitons in the liquid helium~II. It is shown that for some range of temperatures the smallest and most stable roton clusters consist of 13 helium atoms in symmetric vibrational quantum state. A natural model of this roton cluster has a central atom surrounded by a shell of $12$ atoms situated at the vertices of a regular icosahedron~\cite{Krug}; the stability of this configuration is favored by its having the greatest number (six) of nearest neighbors for each atom in a shell. We have found theoretically all parameters defining Landau's roton excitation spectrum, in agreement to high accuracy with experimental data. For example, the roton gap given by the formula $\Delta=\epsilon-\epsilon_0$ coincides to three significant figures (the experimental accuracy) with the experimental value measured by Yarnell {\it et al}~\cite{Yar} from long-wavelength neutron scattering. 
        
\section{references}

\end{document}